\documentstyle[amssymb,aps,preprint,epsf]{revtex}
\begin{document}
%\wideabs{
\title{Systematic low temperature expansion in Ginzburg - Landau model.}
\author{H. C. Kao$^{a,b}$, B. Rosenstein$^{a,c}$ and J. C. Lee$^{c}$}
\address{$^{a}$Academia Sinica, Institute of Physics, Taipei, Taiwan, R.O.C.\\
$^{b}$ Tamkang University, Department of Physics, Tamsui, Taiwan, R.O.C.\\
$^{c}$National Center for Theoretical Studies and Electrophysics Department,%
\\
National Chiao Tung University, Hsinchu, Taiwan, R.O.C.}
\date{August 10, 1999 }
\maketitle

\begin{abstract}
Consistent perturbation theory for thermodynamical quantities in strongly
type II superconductors in magnetic field at low temperatures is developed.
It is complementary to the existing expansion valid at high temperatures.
Magnetization and specific heat are calculated to two loop order and compare
well with existing Monte Carlo simulations, other theories and experiments.
\end{abstract}

\pacs{74.25.Bt, 74.20. De, 74.40. k}
%}
\section{\protect\bigskip Introduction}

Thermal fluctuations play much larger role in high $T_{c}$ superconductors
than in the low temperature ones because the Ginzburg parameter $Gi$
characterizing fluctuations is much larger. In the presence of magnetic
field the importance of fluctuations in high $T_{c}$ superconductors is
further enhanced. First of all, strong magnetic field effectively confines
long wavelength fluctuations in direction perpendicular to the field so that
the dimensionality of the fluctuations are reduced by two \cite{Nelson}.
Moreover highly anisotropic BSCCO type layered superconductors are basically
two dimensional and one expects further increase in the importance of
fluctuations. Under these circumstances fluctuations influence various
physical properties and even lead to new observable qualitative phenomena
like vortex lattice melting into vortex liquid far below the mean field
phase transition line. It is quite straightforward to systematically account
for the fluctuations effect on magnetization, specific heat or conductivity
perturbatively above the mean field transition line using Ginzburg - Landau
description \cite{Tinkham}. However it proved to be extremely difficult to
develop a quantitative theory in the most interesting region below this
line, even neglecting fluctuation of magnetic field and within the lowest
Landau level (LLL) approximation.

To approach the region below the mean field transition line $T<T_{mf}(H)$
Thouless \cite{Thouless} proposed a perturbative approach around homogeneous
(liquid) state in which all the ''bubble'' diagrams are resummed. The series
provides accurate results at high temperatures, but become inapplicable for
LLL dimensionless temperature $a_{T}$ $\thicksim $ $(T-T_{mf}(H))/(TH)^{1/2}$
in $2D$ smaller than $2$ (see the lines H4-6 on Fig.3 for the case of $D=2$
which represent successive approximants (the corresponding three dimensional
plots appear in \cite{Rosenstein}). Generally, attempts to extend the theory
to lower temperature by Pade extrapolation were not successful \cite{Moore1}.

Alternatively, more direct approach to low temperature fluctuation physics would be to
start from the Abrikosov solution at zero temperature and then take into account
perturbatively deviations from this inhomogeneous solution. Experimentally it is
reasonable since, for example, specific heat at low temperatures is a smooth function
and the fluctuation contribution is quite small. This contrasts sharply with
theoretical expectations. Long time ago Eilenberger calculated spectrum of harmonic
excitations of the triangular vortex lattice \cite{Eilenberger}. Subsequently Maki and
Takayama \cite{Maki} noted that the gapless mode is softer than the usual Goldstone
mode expected as a result of spontaneous breaking of translational
invariance. The propagator for the phase excitations behaves as $%
1/(k^{2})^{2}$ in $2D$. This unexpected additional softness not only enhances the
contribution of fluctuations at leading order but also leads to disastrous infrared
divergencies apparently at higher orders. As a result, whether the perturbation theory
around the vortex state is reliable becomes doubtful. For example the contributions to
free energy depicted on Fig. 2a and 2d are respectively $L^{4}$ and $\log ^{2}(L)$
divergent ($L$ being an IR cutoff) in $3D$ and the divergencies get worse at higher
orders . In $2D$ the situation is aggravated: the diagrams diverge as $L^{8}$ and
$L^{4}.$ Also qualitatively the lower critical dimensionality for melting of the
Abrikosov lattice is $D_{c}=3$ and consequently infinite range vortex lattice in clean
materials exists only at $\,\,T=0$. One therefore tends to think that nonperturbative
effects are so important that such a perturbation theory should be abandoned
\cite{Ruggeri}. However a closer look at the diagrams like Fig.2a (see details below)
reveals that in fact one encounters actually only milder divergencies in both $2D$ and
$3D$. This makes the divergencies similar to the so called spurious divergencies in
the theory of critical phenomena with broken continuous symmetry. In such case one can
prove \cite{David} that they exactly cancel at each order provided we are calculating
a symmetric quantity.

In this paper we show that all the IR divergencies in free energy or other
quantities invariant under translations cancel to the two loop order. The $%
3D $ case was briefly reported in \cite{Rosenstein}, here we present in detail the
more complicated $2D$ case. We calculate magnetization and specific heat to this order
and compare the results with existing high temperature expansion, Monte Carlo (MC)
simulation \cite{Nagaosa} of the same system and experiments. Qualitatively physics of
fluctuating $3D$ GL model in magnetic field turns out to be similar to that of $2D$
spin systems (or scalar fields) possessing a continuous symmetry. In particular,
although within perturbation theory the ordered phase (solid) exists only at $T=0$ in
thermodynamic limit, at low temperatures solid with powerwise decay of the
translational order or liquid (exponential decay) differs very little from solid in
most aspects. Therefore, one can effectively use properly modified perturbation theory
to study quantitatively various properties of the vortex liquid phase. Similarly
physics of the $2D$ GL model is analogous to that of the $1D$ scalar theory with, say
$O(2)$ symmetry. This is equivalent to the anharmonic oscillator in quantum mechanics
or the $XY$ spin chain in statistical physics.

The paper is organized as follows. In section II the model is introduced and
the Feynman rules for the loop (low temperature) expansion within the LLL
approximation are set up. The free energy to two loops is calculated using
diagrammatic expansion in section III. It is shown how all the infrared
divergencies cancel exactly at least up to the two loop order. In section IV
we compare the present expansion with the high temperature series, theory of
Tesanovic et al \cite{Tesanovic}, available MC simulations and experiments.
In section V we discuss qualitatively nonperturbative effects using the
analogy with magnetic systems mentioned above. We exemplify this by
calculating perturbatively the ground state energy of the $O(2)$ invariant $%
1D$ chain (equivalent to the quantum mechanical anharmonic oscillator) which
exhibits similar IR divergencies. We argue that although the infinite range
translational order is not present in this system, locally the system looks
like a lattice and the perturbative results are valid up to exponential
corrections. Finally we briefly summarize the results and discuss the
melting transition observed in experiments and some MC simulations,
contributions of higher LL, and fluctuations of magnetic field in section
VII.

\section{Model, mean field solution and the perturbation theory}

\subsection{2D Ginzburg - Landau model}

Our starting point is the GL free energy:
\begin{equation}
F=L_{z}\int d^{2}x\frac{{\hbar }^{2}}{2m}\left| \left( {\bf \nabla }-\frac{%
ie^{\ast }}{\hbar c}{\bf A}\right) \psi \right| ^{2}+a|\psi |^{2}+\frac{%
b^{\prime }}{2}|\psi |^{4}
\end{equation}
Here ${\bf A}=(-By,0) $ describes a nonfluctuating constant magnetic field
in Landau gauge. In strongly type II superconductors ($\kappa \sim 100 $),
far from $H_{c1} $ (this is the range of interest in this paper) magnetic
field is homogeneous to a high degree due to superposition from many
vortices. For simplicity we assume $a=-\alpha ^{\prime }T_{c}(1-t) $, $%
t\equiv T/T_{c} $ although this dependence can be easily modified to better
describe the experimental coherence length.

Throughout most of the paper the following units will be used. Unit of
length is $\xi =\sqrt{{\hbar }^{2}/\left( 2m_{ab}\alpha ^{\prime
}T_{c}\right) } $ and unit of magnetic field is $H_{c2} $, so that the
dimensionless magnetic field is $b\equiv B/H_{c2} $. The dimensionless free
energy in these units is (the order parameter field is rescaled as $\psi
^{2}\rightarrow \frac{2\alpha ^{\prime }T_{c}}{b^{\prime }}\psi ^{2} $):

\begin{equation}  \label{energ1}
\frac{F}{T}=\frac{1}{\omega }\int d^{2}x\left[ \frac{1}{2}|D\psi |^{2}-\frac{%
1-t}{2}|\psi |^{2}+\frac{1}{2}|\psi |^{4}\right] .
\end{equation}
The dimensionless coefficient is
\begin{equation}  \label{omega}
\omega =\sqrt{2Gi}\pi ^{2}t,
\end{equation}
where the Ginzburg number is defined by $Gi\equiv \frac{1}{2}(\frac{32\pi
e^{2}\kappa ^{2}\xi T_{c}\gamma ^{1/2}}{c^{2}h^{2}})^{2} $ and $\gamma
\equiv m_{c}/m_{ab} $ is an anisotropy parameter. The coefficient $\omega $
determines the strength of fluctuations, but is irrelevant as far as mean
field solutions are concerned.

\subsection{Mean field solution near $H_{c2}$}

Define operator ${\cal H}\equiv \frac{1}{2}(-D^{2}-b) $, the free energy
becomes
\begin{equation}  \label{energ2}
\frac{F}{T}=\frac{1}{\omega }\int d^{2}x\left[ \psi ^{\ast }{\cal H}\psi
-a_{h}|\psi |^{2}+\frac{1}{2}|\psi |^{4}\right] .
\end{equation}
Here,
\begin{equation}  \label{ah}
a_{h}\equiv \frac{1-t-b}{2}.
\end{equation}
is the second expansion parameter. If $a_{h} $ is sufficiently small, GL
equations\ can be solved perturbatively (see \cite{Ikeda,Li} for details):

\begin{equation}
\Phi =\sqrt{\frac{a_{h}}{\beta _{A}}}\varphi ({\bf x})+O(a_{h}^{3/2}),
\end{equation}
where $\beta _{A}=1.16 $ and
\begin{equation}
\varphi ({\bf x})=\sqrt{\frac{2\pi b}{\sqrt{\pi }a_{\bigtriangleup }}}\sum
_{l=-\infty }^{\infty }\exp \left\{ i\left[ \frac{\pi l(l-1)}{2}+\frac{2\pi
}{a_{\bigtriangleup }}l\sqrt{b}x\right] -\frac{1}{2}(\sqrt{b}y-\frac{2\pi }{%
a_{\bigtriangleup }}l)^{2}\right\} .
\end{equation}
The lattice spacing is given by $\frac{a_{\bigtriangleup }}{\sqrt{b}} $,
with $a_{\bigtriangleup }=\sqrt{\frac{4\pi }{\sqrt{3}}} $ and $\frac{1}{%
\sqrt{b}} $ the magnetic length in our units. It is normalized to the unit
average $<|\varphi |^{2}>=1 $

\subsection{Fluctuations and Feynman rules}

Within the LLL approximation, which will be used here, the above solution
becomes exact so higher orders in $a_{h}$ do not appear and $\psi $ can be
expanded in a basis of quasimomentum ${\bf k}$ eigenfunctions
\[
\varphi _{{\bf k}}=\sqrt{\frac{2\pi b}{\sqrt{\pi }a_{\bigtriangleup }}}%
\sum_{l=-\infty }^{\infty }\exp \left\{ i\left[ \frac{\pi l(l-1)}{2}+\frac{%
2\pi }{a_{\bigtriangleup }}l(\sqrt{b}x+\frac{k_{y}}{\sqrt{b}})+xk_{x}\right]
-\frac{1}{2}(\sqrt{b}y-{\frac{k_{x}}{\sqrt{b}}}-\frac{2\pi }{%
a_{\bigtriangleup }}l)^{2}\right\} .
\]
around the mean field solution:
\begin{equation}
\psi (x)=v\varphi (x)+\int_{B.z.}{\frac{d^{2}k}{\left( 2\pi \right) ^{2}}}%
\varphi _{{\bf k}}(x)c_{{\bf k}}\left( O_{{\bf k}}+iA_{{\bf k}}\right) ,
\label{shift}
\end{equation}
The shift $v$ of the fluctuating field will be generally different from its
mean field value. The integration is over Brillouin zone of the hexagonal
lattice which has an area $2\pi $. $O_{{\bf k}}$ and $A_{{\bf k}}$ are
''real'' fields satisfying $O_{{\bf k}}^{\ast }=O_{{\bf -k}}$, $A_{{\bf k}%
}^{\ast }=A_{{\bf -k}}$. They are somewhat analogous to the acoustic and
optical phonons in usual solids with some peculiarities due to strong
magnetic field \cite{Moore1}. For example the $A$ mode corresponds to shear
of the two dimensional lattice. After we substitute eq.(\ref{shift}) into
the free energy, the quadratic terms in fields define propagators, while the
cubic and quartic terms give rise to interactions. The phase factors $c_{%
{\bf k}}\equiv \sqrt{\frac{\gamma _{{\bf k}}}{2|\gamma _{{\bf k}}|}}$ with $%
\gamma _{{\bf k}}\equiv \int_{x}\varphi _{{\bf k}}^{\ast }(x)\varphi _{-{\bf %
k}}^{\ast }(x)\varphi (x)\varphi (x)$ are introduced in order to diagonalize
the resulting quadratic part of the free energy
\begin{equation}
F_{quad}=\frac{1}{2}\int_{B.z.}{\frac{d^{2}k}{\left( 2\pi \right) ^{2}}}%
\left[ P_{O}^{-1}({\bf k})O_{{\bf k}}^{\ast }O_{{\bf k}}+P_{A}^{-1}({\bf k}%
)A_{{\bf k}}^{\ast }A_{{\bf k}}\right] .
\end{equation}
Here, $P_{O,A}({\bf k})$ are the propagators entering the Feynman diagrams
in Fig. 1a-b and are given by
\begin{equation}
P_{O,A}({\bf k})=M_{O,A}^{-2}({\bf k});\quad M_{O,A}^{-2}({\bf k}%
)=-a_{h}+v^{2}(2\beta _{{\bf k}}\pm |\gamma _{{\bf k}}|),
\end{equation}
with $\beta _{{\bf k}}\equiv \int_{x}\varphi _{{\bf k}}^{\ast }(x)\varphi
^{\ast }(x)\varphi _{{\bf k}}(x)\varphi (x).$

For convenience, let us introduce the function $\lambda ({\bf k}_{1},{\bf k}%
_{2})$:

\begin{eqnarray}
\lambda ({\bf k}_{1},{\bf k}_{2}) &=&\sqrt{\frac{\sqrt{3}}{2}}\exp \left\{ -%
\frac{(k_{1}^{x})^{2}+(k_{2}^{x})^{2}}{2}\right\} \; \\
&\;&\sum_{l,m}(-)^{lm}\exp \left\{ i\frac{2\pi }{a_{\bigtriangleup }}\left[
lk_{1}^{y}+mk_{2}^{y}\right] -\frac{1}{2}\left[ (k_{2}^{x}+\frac{2\pi }{%
a_{\bigtriangleup }}l)^{2}+(k_{1}^{x}+\frac{2\pi }{a_{\bigtriangleup }}m)^{2}%
\right] \right\} .
\end{eqnarray}
Now, various functions encountered so far as well as all the three and four
leg vertices can be expressed in terms of it: for example, $\gamma _{{\bf k}%
}=\lambda (-{\bf k},{\bf k})$, $\beta _{{\bf k}}=\lambda (0,{\bf k})$. \ The
vertices are depicted on Fig.1c-i with following expressions. Three leg
vertices:
\begin{equation}
(c):A_{{\bf k}_{1}}A_{{\bf k}_{2}}A_{-{\bf k}_{1}-{\bf k}_{2}}=-2v\mathop
{\textrm{Im}}\left[ \lambda ({\bf k}_{2},{\bf k}_{1}+{\bf k}_{2})c_{{\bf k}%
_{1}}^{\ast }c_{{\bf k}_{2}}c_{{\bf k}_{1}+{\bf k}_{2}}\right],
\end{equation}

\bigskip
\begin{equation}
(d): A_{{\bf k}_{1}}A_{{\bf k}_{2}}O_{-{\bf k}_{1}-{\bf k}_{2}}=2v\mathop
{\textrm{Re}}\left[ -\lambda ({\bf k}_{1},-{\bf k}_{2})c_{{\bf k}_{1}}c_{%
{\bf k}_{2}}c_{{\bf k}_{1}+{\bf k}_{2}}^{\ast }+2\lambda (-{\bf k}_{1}-{\bf k%
}_{2},-{\bf k}_{2})c_{{\bf k}_{1}}^{\ast }c_{{\bf k}_{2}}c_{{\bf k}_{1}+{\bf %
k}_{2}}\right],
\end{equation}

\begin{equation}
(e): O_{{\bf k}_{1}}O_{{\bf k}_{2}}A_{-{\bf k}_{1}-{\bf k}_{2}}=-2v\mathop
{\textrm{Im}}\left[ -\lambda ({\bf k}_{1},2{\bf k}_{1}+{\bf k}_{2})c_{{\bf k}%
_{1}}^{\ast }c_{{\bf k}_{2}}c_{{\bf k}_{1}+{\bf k}_{2}}+2\lambda ({\bf k}%
_{1},-{\bf k}_{2})c_{{\bf k}_{1}}c_{{\bf k}_{2}}c_{{\bf k}_{1}+{\bf k}%
_{2}}^{\ast }\right],
\end{equation}

\begin{equation}
(f):O_{{\bf k}_{1}}O_{{\bf k}_{2}}O_{-{\bf k}_{1}-{\bf k}_{2}}=2v\mathop
{\textrm{Re}}\left[ \lambda ({\bf k}_{2},{\bf k}_{1}+{\bf k}_{2})c_{{\bf k}%
_{1}}^{\ast }c_{{\bf k}_{2}}c_{{\bf k}_{1}+{\bf k}_{2}}\right].
\end{equation}

Four leg vertices
\begin{equation}
(g): A_{{\bf k}_{1}}A_{{\bf k}_{2}}A_{{\bf k}_{3}}A_{-{\bf k}_{1}-{\bf k}%
_{2}-{\bf k}_{3}}=\frac{1}{2}\mathop {\textrm{Re}}[\lambda ({\bf k}_{1}+{\bf
k}_{3},{\bf k}_{2}+{\bf k}_{3})c_{{\bf k}_{1}}^{\ast }c_{{\bf k}_{2}}^{\ast }c_{_{{\bf
k}_{3}}}c_{{\bf k}_{1}+{\bf k}_{2}+{\bf k}_{3}}],
\end{equation}

\bigskip

\begin{eqnarray}
(h):A_{{\bf k}_{1}}A_{{\bf k}_{2}}O_{{\bf k}_{3}}O_{-{\bf k}_{1}-{\bf k}_{2}-%
{\bf k}_{3}}=\mathop {\textrm{Re}} &[&-\lambda ({\bf k}_{1}+{\bf k}_{3},{\bf %
k}_{2}+{\bf k}_{3})c_{{\bf k}_{1}}^{\ast }c_{{\bf k}_{2}}^{\ast }c_{_{{\bf k}%
_{3}}}c_{{\bf k}_{1}+{\bf k}_{2}+{\bf k}_{3}} \\
&\;&+2\lambda ({\bf k}_{1}+{\bf k}_{3},{\bf k}_{1}+{\bf k}_{2})c_{{\bf k}%
_{1}}c_{{\bf k}_{2}}^{\ast }c_{_{{\bf k}_{3}}}^{\ast }c_{{\bf k}_{1}+{\bf k}%
_{2}+{\bf k}_{3}}],
\end{eqnarray}

\bigskip

\begin{equation}
(i):O_{{\bf k}_{1}}O_{{\bf k}_{2}}O_{{\bf k}_{3}}O_{-{\bf k}_{1}-{\bf k}_{2}-%
{\bf k}_{3}}=\frac{1}{2}\mathop {\textrm{Re}}[\lambda ({\bf k}_{1}+{\bf k}_{3},{\bf
k}_{2}+{\bf k}_{3})c_{{\bf k}_{1}}^{\ast }c_{{\bf k}_{2}}^{\ast }c_{_{{\bf
k}_{3}}}c_{{\bf k}_{1}+{\bf k}_{2}+{\bf k}_{3}}].
\end{equation}
\bigskip

\section{Cancellation of infrared divergencies in loop expansion}

\subsection{One loop free energy, ''field shift'' and destruction of the
infinite range translation order by fluctuations}

If the fluctuations were absent the expectation value $v_{0}^{2}=\frac{a_{h}%
}{\beta _{A}}$ would minimize $G_{0}=-a_{h}v^{2}+\frac{1}{2}\beta _{A}v^{4}$%
. The one loop contribution to the free energy is
\begin{equation}
G_{1}=\frac{1}{2}\frac{1}{(2\pi )^{2}}\int_{{\bf k}}\left\{ \log [M_{O}^{2}(%
{\bf k})]+\log [M_{A}^{2}({\bf k})]\right\}
\end{equation}
To this order the free energy which is a symmetric quantity is convergent.
However the expectation value of the field which is not a symmetric quantity
is divergent. Minimizing $G_{0}+G_{1}$ with respect to $v$ would lead to the
following correction to the vacuum expectation value:
\begin{equation}
v_{1}^{2}=\frac{1}{2}\frac{1}{(2\pi )^{2}}\int_{{\bf k}}\left\{ \frac{%
[2\beta _{{\bf k}}+|\gamma _{{\bf k}}|]}{M_{O}^{2}({\bf k})}+\frac{[2\beta _{%
{\bf k}}-|\gamma _{{\bf k}}|]}{M_{A}^{2}({\bf k})}\right\} .  \label{corrvev}
\end{equation}

Due to additional softness of the $A$ mode the above integral diverges in
the infrared region. This means that the fluctuations destroy the
inhomogeneous ground state, namely the state with lowest energy is a
homogeneous liquid in accord with the fact that the lower critical dimension
for the present model is $D=3$ \cite{Rosenstein}. It however does not
necessarily means that perturbation theory starting from an ordered ground
state is inapplicable. The way to proceed in such situations have been found
while considering simpler models like $1D$ $\varphi ^{4}$ model $F=\frac{1}{2%
}(\triangledown \varphi_{a})^{2}+V(\varphi_a ^{2})$, $a=1,2$ discussed in
detail in section V (see also \cite{Jevicki}).

\subsection{Two loop contributions to the free energy}

To the two loop order one gets several classes of diagrams (see Fig.2): the
setting-sun ($AAA,AAO,AOO,OOO$), double bubble ($AA,AO,OO$), and the ''correction
term'' ($AA,AA,OO$), which arises due to correction in the value of $v$ from
eq.(\ref{corrvev}) . All of them can be expressed explicitly in terms of the
function $\lambda ({\bf k}_{1},{\bf k}_{2})$ and $d_{{\bf k}}\equiv 2c_{{\bf %
k}}^{2}.$

{\it 1. Setting sun diagrams.} The setting-sun diagrams are shown in Fig.2a-d. The
$AAA$ diagram is naively the most divergent one among them.
\begin{eqnarray}
a:\; AAA &=&\frac{-v^{2}}{8\left( 2\pi \right) ^{2}}\int_{{\bf p}}\int_{{\bf %
q}}\frac{I_{ssAAA}({\bf p},{\bf q})}{M_{A}^{2}({\bf p})M_{A}^{2}({\bf q}%
)M_{A}^{2}({\bf p}+{\bf q})};  \label{diag} \\
I_{ssAAA}({\bf p},{\bf q}) &\equiv &|\lambda ({\bf p},-{\bf q})|^{2}-\lambda
^{2}({\bf p},-{\bf q})d_{{\bf p}}d_{{\bf q}}d_{{\bf p}+{\bf q}}^{\ast } +
\nonumber \\
&&2\lambda ({\bf p},-{\bf q})\lambda ({\bf p},-{\bf p}-{\bf q})d_{{\bf q}}d_{%
{\bf p}+{\bf q}}^{\ast }-2\lambda ({\bf p},-{\bf q})\lambda ^{\ast }({\bf p}%
,-{\bf p}-{\bf q})d_{{\bf p}}+c.c.  \nonumber
\end{eqnarray}
\begin{eqnarray}
b:\; AAO &=& \frac{-v^{2}}{8\left( 2\pi \right) ^{2}}\int _{{\bf q}}\int _{%
{\bf p}} \frac{I_{ssAAO}({\bf p},{\bf q})}{M^2_{A}({\bf p})M^2_{A}({\bf q}%
)M^2_{O}({\bf p}+{\bf q})}; \\
I_{ssAAO}({\bf p},{\bf q}) & \equiv & |\lambda ({\bf p},-{\bf q}%
)|^{2}+\lambda ({\bf p},-{\bf q})^{2}d_{{\bf p}}d_{{\bf q}}d_{{\bf p}+{\bf q}%
}^{*}+2|\lambda ({\bf p}+{\bf q},{\bf p})|^{2}+2\lambda ({\bf p}+{\bf q},%
{\bf p})^{2}d_{{\bf p}}d_{{\bf q}}d_{{\bf p}+{\bf q}} -  \nonumber \\
& & 4\lambda ({\bf p},-{\bf q})\lambda ({\bf p}+{\bf q},{\bf p})d_{{\bf p}}
-4\lambda ({\bf p},-{\bf q})\lambda ^{*}({\bf p}+{\bf q},{\bf p})d_{{\bf q}%
}d_{{\bf p}+ {\bf q}}^{*}+  \nonumber \\
& & 4\lambda ({\bf p}+{\bf q},{\bf p})\lambda ^{*}({\bf p}+{\bf q},{\bf q}%
)d_{{\bf p}}d_{{\bf q}}+4\lambda ({\bf p}+{\bf q},{\bf p})\lambda ({\bf p}+%
{\bf q},{\bf q})d_{{\bf p}+{\bf q}}+c.c.  \nonumber
\end{eqnarray}
\begin{eqnarray}
c:\; AOO &=& \frac{-v^{2}}{8\left( 2\pi \right) ^{2}}\int _{{\bf q}}\int _{%
{\bf p}} \frac{I_{ssAOO}({\bf q},{\bf p})}{M^2{O}({\bf p})M^2{O}({\bf q})M^2{%
A}({\bf p}+{\bf q})}; \\
I_{ssAOO}({\bf p},{\bf q}) & \equiv & |\lambda ({\bf p},-{\bf q}%
)|^{2}-\lambda ({\bf p},-{\bf q})^{2}d_{{\bf p}}d_{{\bf q}}d_{{\bf p}+{\bf q}%
}^{*}+2|\lambda ({\bf p}+{\bf q},{\bf p})|^{2}-2\lambda ({\bf p}+{\bf q},%
{\bf p})^{2}d_{{\bf p}}d_{{\bf q}}d_{{\bf p}+{\bf q}} +  \nonumber \\
& & 4\lambda ({\bf p},-{\bf q})\lambda ({\bf p}+{\bf q},{\bf p})d_{{\bf p}%
}-4\lambda ({\bf p},-{\bf q})\lambda ^{*}({\bf p}+{\bf q},{\bf p})d_{{\bf q}%
}d_{{\bf p}+{\bf q}}^{*}+  \nonumber \\
& & 4\lambda ({\bf p}+{\bf q},{\bf p})\lambda ^{*}({\bf p}+{\bf q},{\bf q}%
)d_{{\bf p}}d_{{\bf q}}-4\lambda ({\bf p}+{\bf q},{\bf p})\lambda ({\bf p}+%
{\bf q},{\bf q})d_{{\bf p}+{\bf q}}+c.c.  \nonumber
\end{eqnarray}
\begin{eqnarray}
d:\; OOO &=& \frac{-v^{2}}{8\left( 2\pi \right) ^{2}}\int _{{\bf q}}\int _{%
{\bf p}} \frac{I_{ssOOO}({\bf p},{\bf q})}{M^2{O}({\bf p})M^2{O}({\bf q})M^2{%
O}({\bf p}+{\bf q})}; \\
I_{ssOOO}({\bf p},{\bf q}) & \equiv & |\lambda ({\bf p},-{\bf q}%
)|^{2}+\lambda ({\bf p},-{\bf q})^{2}d_{{\bf p}}d_{{\bf q}}d_{{\bf p}+{\bf q}%
}^{*}+  \nonumber \\
& & 2\lambda ({\bf p},-{\bf q})\lambda ({\bf p,-p-q})d_{{\bf q}}d_{{\bf p}+%
{\bf q}}^{*}+2\lambda ({\bf p},-{\bf q})\lambda ^{*}({\bf p,-p-q})d_{{\bf p}%
}+c.c.  \nonumber
\end{eqnarray}

{\it 2. Double bubble diagrams.} The bubble diagram are shown in Fig.2e-g.
\begin{eqnarray}
e:\; AA &=& \frac{1}{8\left( 2\pi \right) ^{2}}\int _{{\bf p}}\int _{{\bf q}}%
\frac{I_{bbAA}({\bf p},{\bf q})}{M_{A}^{2}({\bf p})M_{A}^{2}({\bf q})}; \\
I_{bbAA}({\bf p},{\bf q}) & \equiv & \lambda (-{\bf p}+{\bf q},{\bf p}+{\bf q%
})d_{{\bf p}}d_{{\bf q}}+2\beta _{{\bf p-q}}+c.c.  \nonumber
\end{eqnarray}
\begin{eqnarray}
f:\; AO &=& \frac{2}{8\left( 2\pi \right) ^{2}}\int _{{\bf q}}\int _{{\bf p}%
} \frac{I_{bbAO}({\bf p},{\bf q})}{M^2{A}({\bf p})M^2{O}({\bf q})}; \\
I_{bbAO}({\bf p},{\bf q}) & \equiv & -\lambda (-{\bf p}+{\bf q},{\bf p}+{\bf %
q})d_{{\bf p}}d_{{\bf q}}+2\beta _{{\bf p-q}}+c.c.  \nonumber
\end{eqnarray}
\begin{eqnarray}
g:\; OO &=& \frac{1}{8\left( 2\pi \right) ^{2}}\int _{{\bf q}}\int _{{\bf p}%
} \frac{I_{bbOO}({\bf p},{\bf q})}{M^2{O}({\bf p})M^2{O}({\bf q})}; \\
I_{bbOO}({\bf p},{\bf q}) & \equiv & \lambda (-{\bf p}+{\bf q},{\bf p}+{\bf q%
})d_{{\bf p}}d_{{\bf q}}+2\beta _{{\bf p-q}}+c.c.  \nonumber
\end{eqnarray}

\bigskip {}{\it 3. Shift correction terms.} The correction terms are given
by
\begin{eqnarray}
AA &=&\frac{-1}{8\left( 2\pi \right) ^{2}}\int_{{\bf p}}\int_{{\bf q}}\frac{%
I_{crAA}({\bf p},{\bf q})}{M_{A}^{2}({\bf p})M_{A}^{2}({\bf q})}; \\
I_{crAA}({\bf p},{\bf q}) &\equiv &(2\beta _{{\bf p}}-|\gamma _{{\bf p}%
}|)(2\beta _{{\bf q}}-|\gamma _{{\bf q}}|).  \nonumber
\end{eqnarray}
\begin{eqnarray}
AO &=& \frac{-2}{8\left( 2\pi \right) ^{2}}\int _{{\bf q}}\int _{{\bf p}}
\frac{I_{ccAO}({\bf p},{\bf q})}{M^2_{A}({\bf p})M^2_{O}({\bf q})}; \\
I_{crAO}({\bf p},{\bf q}) & \equiv & (2\beta _{{\bf p}}-|\gamma _{{\bf p}%
}|)(2\beta _{{\bf q}}+|\gamma _{{\bf q}}|).  \nonumber
\end{eqnarray}
\begin{eqnarray}
OO &=& \frac{-1}{8\left( 2\pi \right) ^{2}}\int _{{\bf q}}\int _{{\bf p}}
\frac{I_{crOO}({\bf p},{\bf q})}{M^2_{O}({\bf p})M^2_{O}({\bf q})}; \\
I_{crOO}({\bf p},{\bf q}) & \equiv & (2\beta _{{\bf p}}+|\gamma _{{\bf p}%
}|)(2\beta _{{\bf q}}+|\gamma _{{\bf q}}|).  \nonumber
\end{eqnarray}

\subsection{Cancellations of IR divergence within diagrams}

To analyze the IR divergence, one need to expand the propagators and
vertices around small quasimomentum. Using the explicit expansion for $%
\lambda({\bf k}_1, {\bf k}_2)$ given in Appendix A, one can in turn find
those for $\gamma _{{\bf k}}$, $\beta _{{\bf k}}$, and $d_{{\bf k}}$. It
turns out that the constant and $k^2$ terms in $M_{A}^{2}({\bf k})$ vanish,
so that the (only) leading quartic term is $\;-\frac{1}{4}a_{h}\frac{\beta
_{22}}{\beta _{A}}|{\bf k}|^{4}$ and $M_{O}^{2}({\bf k})= a_{h} \left[2 - |%
{\bf k}|^{2} + (\frac{1}{4}-\frac{1}{4}\frac{\beta _{22}}{\beta _{A}})|{\bf k%
}|^{4}\right]$. Here,
\begin{equation}
\beta _{st}\equiv \sqrt{\frac{\sqrt{3}}{2}}\left( \frac{2\pi }{%
a_{\bigtriangleup }}\right) ^{s+t}\sum_{l,m}\,(-)^{lm}\,l^{s}m^{t}\exp \left[
-\frac{(2\pi )^{2}}{2a_{\bigtriangleup }^{2}}(l^{2}+m^{2})\right].
\label{beta}
\end{equation}
\bigskip As a result, the leading divergence $\sim \int_{{\bf p}}\int_{{\bf q%
}}\frac{I_{ssAAA}({\bf p},{\bf q})}{|{\bf p}|^{4}|{\bf q}|^{4}\left| {\bf q}+%
{\bf p}\right| ^{4}}$ is determined by the asymptotics of $I_{ssAAA}({\bf p},%
{\bf q})$ as both ${\bf p}$ and ${\bf q}$ approach zero. If $I_{ssAAA}\sim 1$%
, it would diverge as $L^{8}$. However the vertex is ''supersoft'' at small
quasimomenta so that the divergence is milder than expected. For example,
the $A_{{\bf k}_{1}}A_{{\bf k}_{2}}A_{-{\bf k}_{1}-{\bf k}_{2}}$ vertex
expansion at small momenta starts with $%
(k_{1}^{x}k_{2}^{y}+k_{1}^{y}k_{2}^{x}).$ Therefore the leading divergence
is ''just'' $L^{4}$ . Expanding $I_{ssAAA}({\bf p},{\bf q})$ around ${\bf p}=%
{\bf q}=0$, we see that actually $I_{ssAAA}({\bf p},{\bf q})\propto O(p^{8}):
$%
\[
I_{ssAAA}({\bf p},{\bf q})=\left(\frac{1}{4}\beta _{00}- \beta
_{22}\right) ^{2}\left[p^2 q^2 - (p\cdot q)^2 \right]%
(p^{2}-q^{2})(p^{2}+q^{2}+4p\cdot q).
\]
As a matter of fact, it even becomes $O(p^{10})$ after we symmetrize it with
respect to ${\bf p}\leftrightarrow {\bf q},$ and the diagram is actually
finite. Similarly an apparent logarithmic divergence in setting sun AOO is
nonexistent.

\subsection{Cancellation between different diagrams}

After the apriori most divergent diagram turned out to be convergent we look
for milder IR divergencies in other diagrams. The remaining most divergent
terms appear in contributions $ss_{AAO},bb_{AA}$ and $cr_{AA},$ and come
from the quasimomentum independent terms in the numerator of the integrands:
\[
I_{ssAAO}=4\beta _{00}^{2},I_{bbAA}=3\beta _{00},I_{crAA}=\beta _{00},
\]
respectively. Although they are $L^{4}$ divergent by themselves, their sum with
appropriate coefficients $-(2\beta_{00})^{-1}, 1$ and $-1$ cancels. The order $L^{2}$
divergencies come from the following integrands:
\begin{eqnarray*}
I_{ssAAO} &=&-4\beta _{00}^{2}(p^{2}+q^{2}+2p\cdot q), \\ I_{bbAA} &=&\beta
_{00}(-p^{2}-q^{2}+p\cdot q).
\end{eqnarray*}
Expanding $M^2_{O}({\bf p}+{\bf q})$ in the denominator to the second order in
quasimomenta, we see they cancel each other after
symmetrization with respect to ${\bf p}\leftrightarrow {\bf q}$ and $%
{\bf p}\leftrightarrow -{\bf p}.$ Finally, there are five are $\log (L)$
divergent terms:
\begin{eqnarray*}
I_{ssAAO} &=&\frac{1}{4}\beta _{00}^{2}\left[ 9p^{4}+7q^{4}+ (p\cdot q)
(36p^{2}+28q^{2})+12p^{2}q^{2}+36\left( p\cdot q\right) ^{2}\right]- \\ &&2\beta
_{00}\beta _{22}\left[2(p\cdot q)(p^{2}+q^{2})-p^{2}q^{2}+6\left( p\cdot q\right)
^{2}\right] , \\ I_{bbAA} &=&\frac{1}{4}\beta _{00}\left[ p^{4}+q^{4}-(2p\cdot
q)(p^{2}+q^{2})+6(p\cdot q) ^{2}\right] - \\ &&\frac{1}{4}\beta _{22}\left[
p^{4}+q^{4}-4\left( p\cdot q\right) (p^{2}+q^{2})-6p^{2}q^{2}+20\left( p\cdot q\right)
^{2}\right] , \\ I_{bbAO} &=&\beta _{00}, \\ I_{crAA} &=&-\frac{1}{4}\beta _{22}\left[
p^{4}+q^{4}\right],
\\ I_{crAO} &=&3\beta _{00}.
\end{eqnarray*}
By symmetrizing the sum of all the five terms, we see the final result is
indeed free of IR divergence.

\bigskip

\section{Comparison of results with other theories\ and experiments}

\subsection{Comparison with high temperature expansion}

The same theory has been studied by various analytical and numerical
methods. To compare our results with those obtained using other methods, let
us restore the original units. The Gibbs free energy to two loops (finite
parts of the integrals were calculated numerically) is

\begin{eqnarray}
G &=&\frac{eBk_{B}T}{L_{z}\pi \hbar c}g;  \nonumber \\
g &=&-\frac{1}{2\beta _{A}}a_{T}^{2}+\frac{1}{2\pi }\log (|a_{T}|)+c\frac{1}{%
a_{T}^{2}},  \label{g2D}
\end{eqnarray}
where numerical values of the coefficient is $c=-5.2$. Dimensionless entropy
(the LLL scaled magnetization) is:
\begin{equation}
s=-\frac{dg}{da_{T}}=\left( \frac{L_{z}\pi c^{3}m_{ab}^{2}b^{\prime }}{\hbar
e^{5}k_{B}}\right) ^{1/3}\frac{M}{(TB)^{1/2}}=\frac{1}{\beta _{A}}a_{T}-%
\frac{1}{2\pi }\frac{1}{a_{T}}+2c\frac{1}{a_{T}^{3}},  \label{s2D}
\end{equation}
and specific heat normalized to the mean field value is
\begin{equation}
\frac{1}{\beta _{A}}\frac{C}{\Delta C}=-\frac{d^{2}g}{da_{T}^{2}}=\frac{1}{%
\beta _{A}}+\frac{1}{2\pi }\frac{1}{a_{T}^{2}}-6c\frac{1}{a_{T}^{4}}.
\label{c2D}
\end{equation}
We first compare the results with those of the high temperature expansion
\cite{Thouless}. These series are known now to the 12th order in $x$ $\cite
{Hikami}$ Successive partial sums for specific heat at low temperature are
plotted on Fig.3 (dashed lines) together with several orders of the high
temperature expansion. Low temperature expansion indicates that the specific
heat ratio grows with $a_{T}$. On the other hand, the high temperature
expansion clearly shows that it drops out fast above $a_{T}=0$. This means
that there is a maximum in between which is consistent with most experiments
and Monte Carlo simulations, see Fig. 4. Whether there is a melting phase
transition either first order or continuous (in $2D$ it is necessarily of
the Kosterlitz - Thouless variety) cannot be determined from series alone.
Both series expansions have a finite radius of convergence, but this fact
alone is not enough to decide that the singularity is at real temperature
(it can be located in the complex plane as for example in the $1D$ Ising
model). The low temperature series are too short to estimate the radius of
convergence. Naively comparing the second coefficient in specific heat eq.(%
\ref{c2D}) to the third one obtains an estimate $a_{T}=-\sqrt{12\pi c_{3}}=-8
$. Phenomenologically first order melting occurs around $a_{T}=-10$.
Extensive analysis of the high temperature series has been made in \cite
{Hikami,Wilkin1}.

\subsection{Comparison with MC, experiments and other theoretical results}

Low temperature results for free energy and magnetization agree well with
available numerical simulations and experiments. However specific heat
comparison is the most sensitive (second derivative). We therefore present
here only the specific heat comparison. For $a_{T}$ $<-5$ the specific heat
results on Fig. 4 are in accord qualitatively with experiments of \cite
{Kapitulnik}, and MC simulations of \cite{Nagaosa}. The same data were
fitted by the theory of Tesanovic and coworkers \cite{Tesanovic}, We can
calculate the coefficients of the low energy expansions from their theory
and we get:
\[
\frac{1}{\beta _{A}}\frac{C}{\Delta C}=\frac{1}{\beta _{A}}-2\frac{1}{%
a_{T}^{2}}+12\beta _{A}\frac{1}{a_{T}^{4}}
\]
compared to our eq.(\ref{c2D}). Even the sign of the $1/a_{T}^{2}$
contribution is different. This theory is only an approximate one and
perhaps some modifications are required in the low temperature limit.

\section{Understanding cancellations of infrared divergencies in a simple
model. Nonperturbative effects}

\subsection{Toy model}

The dramatic cancellation of all the severe IR divergencies in the GL LLL
model up to $L^{8}$ at the two loop level looks a bit mysterious. Although
in critical phenomena the phenomenon of cancellation of ''spurious''
divergencies due to Goldstone bosons is well known \cite{Lawrie}, here it
occurs under rather extreme circumstances. The theory is below its lower
critical dimensionality. To better understand what is involved in these
cancellations we investigated a model which has similar symmetry properties,
but is much simpler. As was mentioned in the introduction the physics of the
$D$ dimensional GL theory is very reminiscent of that of the $D-1$
dimensional scalar theory with two fields possessing a continuous $O(2)$
symmetry, see Fig. 5,
\[
F=\frac{1}{2}(\stackrel{\cdot }{\varphi }_{a})^{2}+a_{T}\varphi_a^{2}+\frac{1%
}{2}\left( \varphi_a ^{2}\right) ^{2}
\]
where the dot denotes the derivative in the only dimension considered as
''time'' and we are interested in the spontaneously broken symmetry case $%
a_{T}<0$.

This model is equivalent to quantum mechanics of the two dimensional
anharmonic oscillator. Of course one can solve this model nonperturbatively
(albeit using numerical solution of the differential equation, we are not
aware of the closed form of the ground state energy). Obviously the result
is IR finite (bounded from below by the classical energy and from above by
variational Gaussian energy) \cite{Stevenson}.

\subsection{Perturbation theory}

It is important to trace the origin of the IR divergencies in the
intermediate steps of the perturbative calculation. Although the calculation
has been done using Feynman diagrams (steepest descent approximation of the
path integral), like in GL\ model above, it is useful to start from the
standard time independent perturbation theory. Here we first have to decide
what is the main part $K$ and what will be a perturbation $V$. The
perturbative vacuum in which $<\varphi_{a}>=v_{a}$ is degenerate and we have
many choices of the ''unperturbed part. One of them corresponding to the
choice
\begin{eqnarray*}
v_{a} &=&(\sqrt{-a_{T}},0)\equiv (v,0), \\
O &\equiv &\varphi_{1}-v,\text{ \ \ \ }A\equiv \varphi_{2}
\end{eqnarray*}
is:
\begin{eqnarray*}
H &=&-\frac{a_{T}^{2}}{2}+K+V, \\
K &=&\frac{1}{2}(\pi _{O}{}^{2}+\pi _{A}^{2}{})+2v^{2}O^{2}+\frac{1}{2L^{2}}%
A^{2}, \\
V &=&2v(O^{3}+OA^{2})+\frac{1}{2}\left( O^{2}+A^{2}\right) ^{2}
\end{eqnarray*}
where $\pi _{O}$ and $\pi _{A}$ are conjugate momenta of the fields $O$ and $%
A$ respectively. The constant in $H$ is the classical energy. Since any of
these ground states are nonnormalizable, see Fig.5, the IR cutoff $L$ was
introduced into $K$. It also removes the vacuum degeneracy. With this cutoff
the unperturbed wave function is:
\[
\Psi _{0,0}(\varphi _{1},\varphi _{2})=\left( \frac{2v}{\pi ^{2}L}\right)
^{1/4}\exp \left[ -\frac{1}{2L}\varphi _{2}^{2}-v(\varphi _{1}-v)^{2}\right]
.
\]

Zero point energy $\left\langle \Psi _{0,0}|V|\Psi _{0,0}\right\rangle $
corresponds in the time dependent perturbation theory to
\[
F_{1}=\frac{1}{2}\left( Tr\log G_{O}+Tr\log G_{A}\right) =v+O\left( \frac{1}{%
L}\right)
\]
in the time dependent formalism. The leading correction to the ground state
energy is:
\[
F_{2}^{bb}=\left\langle \Psi _{0,0}|V|\Psi _{0,0}\right\rangle =\frac{3L^{2}%
}{8}+\frac{L}{8}\frac{1}{v}+\frac{3}{32}\frac{1}{v^{2}}+O\left( \frac{1}{L}%
\right) .
\]
It is equal to three ''double bubble diagrams of Fig.2. The second order in $%
V$ correction is:
\begin{eqnarray*}
F_{2}^{ss} &=&\sum_{(n1,n2)\neq (0,0)}\frac{|\left\langle \Psi _{0,0}|V|\Psi
_{n1,n2}\right\rangle |^{2}}{E_{0,0}-E_{n1,n2}} \\
&=&\frac{|\left\langle \Psi _{0,0}|V|\Psi _{1,0}\right\rangle |^{2}}{-2v}+%
\frac{|\left\langle \Psi _{0,0}|V|\Psi _{3,0}\right\rangle |^{2}}{-6v}+\frac{%
|\left\langle \Psi _{0,0}|V|\Psi _{1,2}\right\rangle |^{2}}{-2v-\frac{2}{L}}
\\
&=&-\frac{3L^{2}}{8}-\frac{L}{8}\frac{1}{v}-\frac{19}{32}\frac{1}{v^{2}}%
+O\left( \frac{1}{L}\right) .
\end{eqnarray*}
This contribution is the sum of the ''setting sun diagrams'' and correction
terms (some terms which contain higher orders in $1/a_{T}$ were dropped).
Unlike the GL theory there is no $AAA$ setting sun diagram and therefore no $%
L^{4}$ divergence is expected. The leading divergence $L^{2}$ and the
subleading $L$ cancel between $F_{2}^{bb}$ and $F_{2}^{ss}$.

\subsection{ Absence of long range order}

As is well known even discrete symmetry cannot be broken spontaneously in $%
1D.$ This means that when we calculate perturbatively VEV of a quantity
which is not invariant under the symmetry group $O(2)$ it should be IR
divergent \cite{Jevicki}. As an example we calculate expectation value of $%
\varphi _{1}$. To first order correction to $\left\langle \Psi |\varphi
_{1}|\Psi \right\rangle $ arises from the corrected ground state
\begin{eqnarray*}
\Psi &=&\Psi _{0,0}+\sum_{(n1,n2)\neq (0,0)}c_{n1,n2}\Psi _{n1,n2},\  \\
c_{n1,n2} &=&\frac{\left\langle \Psi _{0,0}|V|\Psi _{n1,n2}\right\rangle }{%
E_{0,0}-E_{n1,n2}}, \\
c_{1,0} &=&-\frac{2vL+3}{8v^{3/2}}.
\end{eqnarray*}
The result is
\begin{eqnarray*}
\left\langle \Psi |\varphi _{1}|\Psi \right\rangle &=&v+2c_{1,0}\left\langle
\Psi _{0,0}|O|\Psi _{1,0}\right\rangle \\
&=&\sqrt{-a_{T}}-\frac{1}{\sqrt{-a_{T}}}\frac{L}{4}.
\end{eqnarray*}
It diverges linearly indicating ''dynamical restoration'' of the symmetry.
The sign of the correction indicates that the VEV is reduced. The exact
finite size expression is nonanalytic (like $\sqrt{-a_{T}}\exp \left[ \frac{L%
}{4\sqrt{-a_{T}}}\right] $) and approaches zero.

This model clearly teaches us that although the $O(2)$ symmetry is unbroken,
the perturbation theory starting from the ''broken'' symmetry ground state
not only cures its own IR divergencies problems, but provides an accurate
approximation to any $O(2)$ symmetric quantity. Perturbation theory actually
''knows'' about restoration of the symmetry missing only corrections of the
essential singularity variety.

\section{Conclusion\protect\bigskip{}}

To summarize, it is established up to order of two loops that perturbation
theory around Abrikosov lattice is consistent. All the IR divergencies
cancel due to soft interactions of the soft mode. Perturbative results as
well as interpolation with the high temperature expansion agree very well
with the direct MC simulation and experiments. The theory of Te\v{s}%
anovi\'{c} {\it et al} \cite{Tesanovic} has different low temperature limit
and perhaps should be modified in this region.

Consistency of the perturbation theory rules out a possibility of infinite
lattice at any $T>0$. Let us briefly summarize what it physically means and
under what conditions this result is valid. The fact that $v$ is IR
divergent in both $2D$ (power) and $3D$ (logarithm) means that order
parameter of translational symmetry breaking vanishes. Its correlator at
very large distances approaches zero. This does not necessarily means that
the state is liquid, namely the correlator decays exponentially with certain
correlation length. It might decay only as a power like in $2D$ $XY$ model
\cite{KT} and melt into liquid via either first order or continuous
transition. Assumptions are: no disorder, infinite sample, LLL
approximations, no fluctuations of magnetic field. The third assumption can
be relaxed since one obtains a supersoft $1/k^{4}$ spectrum also after
including higher Landau levels, see \cite{Ikeda,Li}, however including
fluctuations of the magnetic field will probably stabilize the lattice in $3D
$ since the spectrum becomes the usual Goldstone boson $1/k^{2}$.

Author is very grateful to B. Ya. Shapiro, T.K. Lee and D.P. Li for numerous
discussions. Work was supported by grant NSC of Taiwan.

\subsection{Appendix A. Small momentum expansion of Feynman diagrams}

In this Appendix we give formulas for expansion of integrands in powers of
quasimomenta, which is needed to eighth order. The basic quantity which
enter the Feynman diagrams is the function $\lambda $. Using identities
between the coefficients $\beta $ defined in eq.(\ref{beta}): $\beta _{02}=%
\frac{1}{2}\beta _{00},\beta _{04}=\frac{3}{2}\beta _{00}-3\beta _{22},$ we
have
\begin{eqnarray*}
\lambda ({\bf k}_{1},{\bf k}_{2}) &=& \exp\left\{ -\frac{(k_1^x)^2 +
(k_2^x)^2}{2} \right\} \times \\
&& \beta _{00} + \frac{1}{4}\beta _{00}\left[ (k_2^x -ik_1^y)^2 + (k_1^x
-ik_2^y)^2 \right] + \frac{1}{16}\beta_{00} \left[(k_2^x -ik_1^y)^4 + (k_1^x
-ik_2^y)^4 \right] - \\
&& \frac{1}{8}\beta _{22} \left[-(k_2^x -ik_1^y)^2 + (k_1^x -ik_2^y)^2 %
\right]^{2} + \\
&& \frac{1}{720}\beta _{06}[\;\left( k_{2}^{x}-ik_{1}^{y}\right) ^{6}+\left(
k_{1}^{x}-ik_{2}^{y}\right) ^{6}]+ \\
&&\frac{1}{48}\beta _{24}\left( k_{2}^{x}-ik_{1}^{y}\right) ^{2}\left(
k_{1}^{x}-ik_{2}^{y}\right) ^{2}[\left( k_{2}^{x}-ik_{1}^{y}\right)
^{2}+\left( k_{1}^{x}-ik_{2}^{y}\right) ^{2}]+ \\
&&\frac{1}{40320}\beta _{08}[\;\left( k_{2}^{x}-ik_{1}^{y}\right)
^{8}+\left( k_{1}^{x}-ik_{2}^{y}\right) ^{8}]+ \\
&&\frac{1}{1440}\beta _{26}\left( k_{2}^{x}-ik_{1}^{y}\right) ^{2}\left(
k_{1}^{x}-ik_{2}^{y}\right) ^{2}[\left( k_{2}^{x}-ik_{1}^{y}\right)
^{4}+\left( k_{1}^{x}-ik_{2}^{y}\right) ^{4}]+ \\
&& \frac{1}{576}\beta _{44}\left( k_{2}^{x}-ik_{1}^{y}\right) ^{4}\left(
k_{1}^{x}-ik_{2}^{y}\right) ^{4}
\end{eqnarray*}

\newpage
\begin{figure}[tbp]
\epsfysize=2truein \centerline{\epsffile{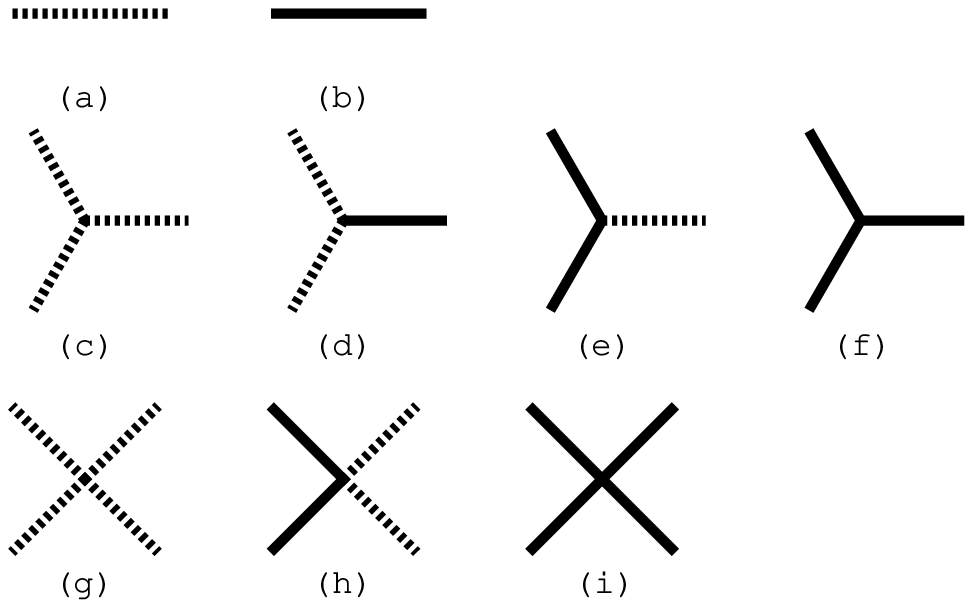}} \caption{All
the relevant propagators and vertices in the theory. Here, dash
and solid lines $\quad $ represent the A and O fields,
respectively.} \label{fig1}
\end{figure}

\begin{figure}[tbp]
\epsfysize=2truein \centerline{\epsffile{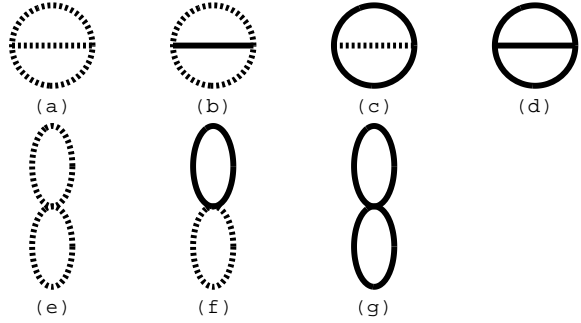}}
\caption{Setting sun and double bubble diagrams that contribute to
the free energy at the two loop level.} \label{fig2}
\end{figure}

\begin{figure}[tbp]
\epsfysize=2truein \centerline{\epsffile{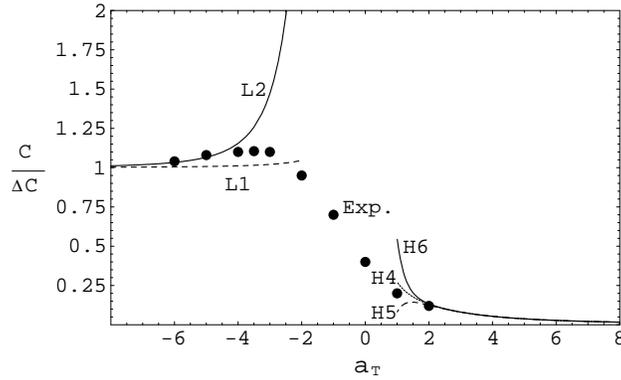}}
\caption{Comparison between low temperature and high temperature
expansion of the scaled specific heat defined in eq.(\ref{c2D}).}
\label{fig3}
\end{figure}

\begin{figure}[tbp]
\epsfysize=2truein \centerline{\epsffile{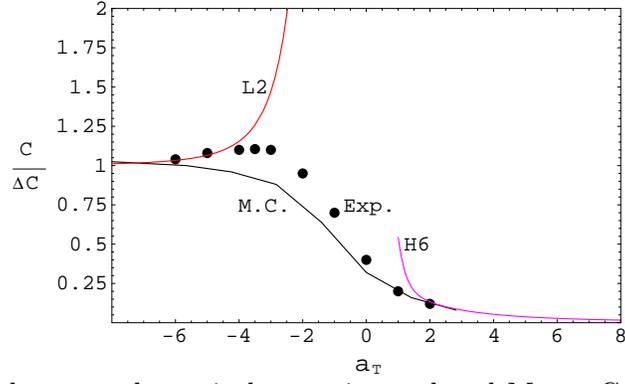}}
\caption{Comparison between theoretical, experimental and Monte
Carlo results of the scaled specific heat.} \label{fig4}
\end{figure}

\begin{figure}[tbp]
\epsfysize=2truein \centerline{\epsffile{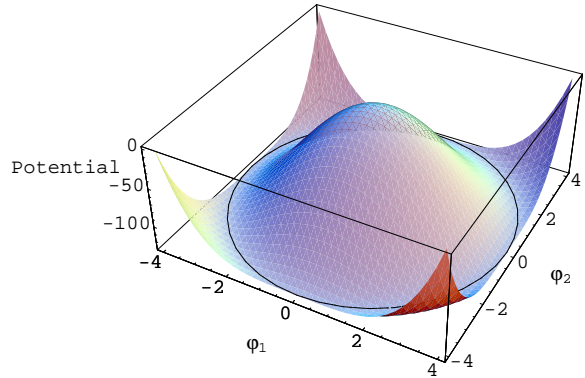}}
\caption{Potential of the anharmonic oscillator. The classical
ground state indicated by the circle is degenerate.} \label{fig5}
\end{figure}

\end{document}